\documentclass[aps,groupaddress]{revtex4}
\usepackage{graphicx}
\usepackage{amssymb}
\usepackage{amsmath}

\begin{document}
\title{Comment on "Deformed Fokker-Planck equation: inhomogeneous medium with a position-dependent mass"}
%

\author{J. {\L}uczka}
\affiliation{Institute of Physics, University of Silesia, 41-500 Chorz{\'o}w, Poland}
\email{jerzy.luczka@us.edu.pl}
\begin{abstract} 
In a recent paper by B. G. da Costa {\it et al.} [Phys. Rev. E 102, 062105(2020)], the phenomenological Langevin equation and the corresponding Fokker-Planck equation for an inhomogeneous medium with a position-dependent particle mass and position-dependent damping coefficient  have  been studied. The aim of this comment is to present a microscopic derivation of the Langevin equation for such a system.  It is not equivalent to that in the commented paper.
\end{abstract}
\maketitle

In Ref. [1] the authors write that the goal of the paper is to present the Fokker-Planck equation for an inhomogeneous medium with a variable diffusion coefficient within the position-dependent mass scenario.  They consider a particle of mass $m(x)$ in a fluid of viscosity coefficient $\lambda(x)$ subjected to an external potential force $F(x)$ and a random force $R(t)$.   They propose  the Langevin equation of the form (see Eq. (46) therein)
\begin{equation} \label{Costa} 
m(x) \ddot{x} + \frac{1}{2} m'(x) \dot{x}^2 = -m(x) \lambda(x) \dot{x} +F(x)+  R(t), 
\end{equation}
where the dot and the prime denotes a  differentiation with respect to time and the coordinate $x$, respectively. 
Next, they write that in the overdamped limit ($\lambda(x) >> \tau^{-1}$ with $\tau$ a coarse-grained timescale) the left-hand side of Eq. (1) vanishes  and  then it reduces to the form 
\begin{equation} \label{Costa2} 
 \dot{x} = \frac{1}{m(x) \lambda(x)} \left[F(x)+  R(t)\right].  
\end{equation}
The authors do not define in a precise way quantities and parameters of the model. In consequence, the reader is confused. If $m(x) =m_0=$ const. and $\lambda(x) =\lambda_0=$ const. then one can  rescale parameters to obtain a consistent description. E.g for the  Stokes frictional force, the first term on the right-hand side of Eq. (1) is presented in a standard form $-\gamma_0 \dot x$, where $\gamma_0$ is the friction coefficient and for consistency one should assume that $\lambda_0=\gamma_0/m_0$. Then one obtains a correct equilibrium state.  However, in the case of $x$-dependence,  from Eq. (2) it follows that 
in the stationary state the corresponding probability distribution $P_s(x)$ for the particle position  depends on its mass $m(x)$ which is not correct. The authors  cite the van Kampen paper on diffusion in inhomogeneous media \cite{kampen}, however  $P_s(x)$ therein does not depend on the particle mass. 

  We want to present an alternative,  more consistent and   microscopic model in terms of 
the Caldeira-Leggett framework. The  system consists of a Brownian particle plus thermostat of an infinite number of harmonic oscillators. The Hamilton function for such a system reads \cite{Leggett,weis}
\begin{equation}
H=\frac{p^2}{2m(x)}+V(x) + \sum_i \left[ \frac{p_i^2}{2m_i} + \frac{m_i
\omega_i^2}{2} \left( q_i - \eta_i g(x)\right)^2 \right], 
\end{equation}
where the coordinate and momentum  $\{x, p\}$ refer to the Brownian particle  of   coordinate-dependent mass $m=m(x)$ subjected to the potential $V(x)$ and $\{q_i, p_i\}$  
  are the coordinate and momentum of the $i$-th heat bath oscillator of mass $m_i$ and the eigenfrequency $\omega_i$. The function $g(x)$ couples the bath modes nonlinearly to the coordinate of the Brownian particle and the parameter $\eta_i$ characterizes the interaction strength of the particle with the $i$-th oscillator.  All coordinate and momentum variables   obey canonical equal-time Poisson-bracket relations. 
 
The next step is to write the Hamilton  equations of motion for all coordinate and momentum variables. 
 For the Brownian particle, the Hamilton equations read 
\begin{align}
\dot{x}  &= \frac{\partial H}{\partial p} = \frac{1}{m(x)} \, p, \label{eq6}\\
\dot{p}  &= -\frac{\partial H}{\partial x} = \frac{m'(x) p^2}{2 m^2(x)} -V'(x) + g'(x) \sum_i c_i \left(q_i - \eta_i g(x) \right), \label{eq7} 
\end{align}
where
\begin{equation}
c_i=\eta_i m_i \omega_i^2. 
\end{equation}

For the thermostat variables, one gets: 
\begin{align}
\dot{q_i}&=\frac{\partial H}{\partial p_i} = \frac{p_i}{m_i}, \label{eq9}\\
\dot{p_i}&=-\frac{\partial H}{\partial q_i} = - m_i \omega_i^2 q_i+c_i g(x).\label{eq10}
\end{align}

What we need in Eq. (\ref{eq7}) is the solution $q_i=q_i(t)$, which can be obtained from Eqs. (\ref{eq9}) and (\ref{eq10}) with the result 
\begin{equation}\label{q solution}
q_i(t)=q_i(0) \cos(\omega_i t) + \frac{p_i(0)}{m_i \omega_i} \sin(\omega_i t) +\frac{c_i}{m_i \omega_i} \int_{0}^{t} \sin[\omega_i(t-s)] g(x(s)) ds.
\end{equation} 

The following step is to integrate by parts the last term in Eqs. (\ref{q solution}) and insert it into Eq. ~(\ref{eq7}) for $p=p(t)$. Using Eq.  (\ref{eq6}), after some algebra, one can obtain an effective equation of motion for the particle coordinate $x(t)$. It is called a generalized Langevin equation 
 and has the form
\begin{align}\label{GLE}
m(x(t))\, {\ddot x}(t) + \frac{1}{2} m'(x(t) ) \, \dot x^2(t) = &-V'(x(t)) 
- g'(x(t)) \, \int_0^t \gamma(t-s) g'(x(s)) \dot{x}(s) \, ds  \nonumber \\
&-\gamma(t) g'(x(t)) g(x(0)) + g'(x(t)) R(t), \quad t>0, 
\end{align}
where  $\gamma(t)$ is a dissipation function (damping or memory kernel) and $R(t)$ denotes the random force, 
\begin{align} 
\gamma(t) &=\sum_i \frac{c_i^2}{m_i \omega_i^2} \cos(\omega_i t), \label{diss}\\
R(t) &=\sum_i c_i \left[q_i(0) \cos(\omega_i t) + \frac{p_i(0)}{m_i \omega_i}
\sin(\omega_i t) \right].
 \label{force} 
\end{align}
In the standard approach, it is assumed that the initial probability distribution of the total system is factorized with an arbitrary probability density for the Brownian particle and the canonical Gaussian distribution for thermostat. In such a case,  $R(t)$ is a Gaussian random force of zero mean obeying the fluctuation-dissipation relation 
\begin{equation} \label{corel}
\langle R(t)R(s) \rangle = k_BT \gamma(t-s),  
\end{equation}
where the brackets denote canonical averaging over the bath modes.

The dynamics of the Brownian particle is therefore described by a stochastic
integro-differential equation for the coordinate  $x(t)$. In the case of a constant mass 
$m(x)=m$,  Eq. (\ref{GLE}) has been intensively applied  in the theory of thermally 
activated rate processes \cite{,weis,Dmit}. When  $m(x)=m$ and $g(x)=x$  it reduces to the standard generalized Langevin equation, namely,  
\begin{align}\label{GLE2}
m{\ddot x}(t)
+\int_0^t \gamma(t-s) \dot{x}(s) \, ds = - V'(x(t)) 
-\gamma(t) x(0)+ R(t), \quad t>0. 
\end{align}
 Now, we can assume the Ohmic dissipation which means that the memory kernel is a Dirac delta,  $\gamma(t)=2\gamma_0 \delta(t)$, where $\gamma_0 >0$.  In  such a case, Eq. (\ref{GLE}) takes the form 

\begin{align}\label{LE}
m(x)\, {\ddot x} + \frac{1}{2} m'(x) \, \dot x^2 = - \gamma_0 [g'(x)]^2 \dot{x} + F(x) +
g'(x) R(t), \quad x=x(t), \quad F(x) = -V'(x), \quad t>0.  
\end{align}
This equation is not equivalent to Eq. (1). First, Eq. (\ref{LE}) is a stochastic equation  with multiplicative noise. Secondly, the deterministic part is  different. In particular, the damping term does not depend on the particle mass. 
The authors of Ref. [1] neglect two terms in the left-hand side of Eq. (1) (calling this procedure as an overdamped limit) and analyse the corresponding Fokker-Planck equation which depends on mass $m(x)$ via the first term in the right-hand side of Eq. (1). If one apply the same procedure to Eq. (\ref{LE}) then the corresponding Fokker-Planck equation takes a different form and does not depend on mass $m(x)$ as it should be. Finally, we want to stress that the overdamped limit has to  be precisely defined by rescaling Eq. (1) and (15), see the critical discussion in Sec. 8 of Ref. \cite{lucz}.

\end{document}